\documentclass[
 reprint,
 amsmath,amssymb,
 aps,
 showkeys,
 superscriptaddress
]{revtex4-2}

\usepackage{graphicx} 
\usepackage{dcolumn} 
\usepackage{bm} 
\usepackage{multirow}
\usepackage{nomencl} 
\usepackage{hyperref} 
\usepackage{caption}
\usepackage{subcaption}
\usepackage{gensymb}
\usepackage{xurl}
\usepackage{diagbox}
\usepackage{threeparttable}
\usepackage{xcolor}
\usepackage{booktabs}
\usepackage{tikz}
\usetikzlibrary{arrows.meta, positioning}
\usepackage{pgfplots}
\usepgfplotslibrary{fillbetween}
\pgfplotsset{compat=1.18}

\makenomenclature

\begin{document}

\preprint{APS/123-QED}

\title{Bounded fuzzy logic control for optimal scheduling of green hydrogen production and revenue maximisation}

\author{Sleiman Farah}
\altaffiliation{sleiman.farah@mymail.unisa.edu.au}
\affiliation{Department of Mechanical and Production Engineering, Aarhus University, Denmark}
\affiliation{Independent researcher, Aarhus, Denmark}

\author{Jens Jakob S{\o}rensen}
\affiliation{{\O}rsted, Kr\ae ftv\ae rksvej 53, 7000 Fredericia}

\author{Kary Fr\"{a}mling}
\affiliation{Ume\r{a} University, Computing Science, 901 87 Ume\r{a}, Sweden}
\affiliation{Aalto University, Department of Industrial Engineering and Management, Maarintie 8, FI-00076 Aalto, Finland}

\author{Matej Simurda}
\affiliation{{\O}rsted, Kr\ae ftv\ae rksvej 53, 7000 Fredericia}

\begin{abstract}

Hydrogen Purchase Agreements (HPAs) guarantee revenue streams that mitigate the financial risks inherent in the long-term production of green hydrogen from renewable energy sources. However, the intermittency of renewable electricity and the availability of parallel revenue opportunities in both the electricity and hydrogen markets complicate the scheduling of green hydrogen production. The scheduling should maximise the total revenue from short-term sales of electricity and hydrogen against the long-term HPA delivery obligations. This challenge is addressed by developing a Bounded Fuzzy Logic Control (BFLC) which determines the daily HPA delivery target based on day-ahead forecasts of electricity and hydrogen prices, as well as wind capacity factors. Subsequently, the daily target is imposed as a constraint in dispatch optimisation which allocates energy and hydrogen flows for each hour of the day.
Revenue comparisons over several years demonstrate that the BFLC achieves total annual revenues within 9\% of optimal revenues that are based on perfect foresight. The BFLC revenues consistently exceed those of steady control, with the largest differences observed under conditions of elevated price levels and variability. The BFLC provides an effective long-term scheduling of green hydrogen production, enabling realistic revenue quantification that mitigates economic risks without overlooking economically viable projects.

\nomenclature{HPA(s)}{Hydrogen purchase agreement(s)}

\end{abstract}

\keywords{green hydrogen; fuzzy logic; hydrogen purchase agreement; renewable energy system; optimal control; revenue maximisation}
\maketitle

\section*{Nomenclature}
\printnomenclature
\nomenclature{$e_t^d$}{Electricity price at hour \( t \) on day \( d \) (€/MWh)}
\nomenclature{$h_t^d$}{Hydrogen price at hour \( t \) on day \( d \) (€/kg)}
\nomenclature{$w_t^d$}{Wind capacity factor at hour \( t \) on day \( d \) (-)}
\nomenclature{$g_{1,t}^d$}{Wind energy at hour \( t \) on day \( d \) (MWh)}
\nomenclature{$g_{2,t}^d$}{Energy to electricity grid at hour \( t \) on day \( d \) (MWh)}
\nomenclature{$g_{3,t}^d$}{Energy to electrolyser at hour \( t \) on day \( d \) (MWh)}
\nomenclature{$m_{1,t}^d$}{Hydrogen mass produced at hour \( t \) on day \( d \) (kg)}
\nomenclature{$m_{2,t}^d$}{Hydrogen mass sent to purchase agreement at hour \( t \) on day \( d \) (kg)}
\nomenclature{$m_{3,t}^d$}{Hydrogen mass sent to market at hour \( t \) on day \( d \) (kg)}

\section{Introduction}
Global energy-related $\text{CO}_2$ emissions reached 36.3~Gt in 2022~\cite{deng2025global}, a level expected to drive substantial increases in global temperatures without immediate and sustained mitigation efforts~\cite{lee2023ipcc, change2022mitigating}. These emissions stem primarily from the continued global dependence on fossil fuels, which dominate energy use across critical sectors such as electricity production, transportation, industry, and space heating~\cite{lee2023ipcc, iea2023co2}.

In the past two decades, substantial progress has been made in the development and deployment of renewable electricity generation technologies, most notably wind and solar photovoltaics. Since 2015, the global installed capacity of these sources has more than doubled, and further expansion is anticipated~\cite{renewables2022, irena2025}. However, despite this growth, the full decarbonisation potential of renewables is limited by their inherent intermittency. Wind and solar generation exhibit variability across multiple timescales---from intra-day fluctuations to seasonal cycles---which poses significant challenges to maintaining a stable and reliable energy supply~\cite{mathiesen2015smart}.

To address this variability, balancing strategies that can match electricity demand with fluctuating renewable generation are required. This balancing can be achieved through a combination of demand-side flexibility and energy storage technologies. Among the available storage options, batteries have emerged as the most mature solution for daily short-duration energy storage. Battery costs have declined significantly over the last decade~\cite{cole2021cost}, making them increasingly viable for commercial and grid-scale applications. Moreover, batteries dominate the global landscape of deployed storage technologies due to their fast response times and integration with existing energy systems~\cite{sivaram2018need}.

Nevertheless, batteries remain unsuitable for decarbonising hard-to-electrify sectors, such as heavy transportation, shipping, and aviation. The key limitation lies in the relatively low energy density (kWh/kg)~\cite{guerra2021beyond}. To overcome the limitations of batteries in the hard-to-electrify sectors, alternative energy carriers are needed. One promising approach involves producing energy-dense fuels from renewable electricity, enabling decarbonisation of the aforementioned sectors. This concept, known as power-to-X, originated in Germany in the early 2010s~\cite{sterner2021power}.

At the core of power-to-X lies power-to-hydrogen, whereby the so-called green hydrogen is produced through electrolysis powered by renewable energy~\cite{oliveira2021green}. This process offers a carbon-free and geographically flexible means of generating hydrogen since electrolysis does not rely on fossil fuel inputs, and water is abundantly available.  Hydrogen's high energy density makes it particularly suitable for applications complementary to batteries where weight constraints are critical~\cite{ball2009future, zeyen2023endogenous}.

Despite its substantial theoretical potential~\cite{neumann2023potential}, the green hydrogen sector in Europe has faced a range of challenges that have led to the downsizing, postponement, or cancellation of numerous power-to-X projects. Only 17\% of the European hydrogen portfolio is expected to materialise by 2023 due to high costs, underdeveloped demand-side markets, and the inability to obtain funding~\cite{westwood2025hydrogen}. As an example {\O}rsted abandoned its e-methanol project in \"{O}rnsk\"{o}ldsvik, Sweden, due to e-Fuel market developing slower than expected~\cite{orsted2024h1}. In Spain, Iberdrola revised its 2030 green hydrogen production from 350,000 to 120,000~tonnes because of challenges with the offtake hydrogen price~\cite{reuters2024iberdrola}.

A central parameter to optimise in the business case for a hydrogen plant is the hydrogen purchase agreement (HPA), which is a long-term purchase contract between a hydrogen producer and a buyer (offtaker), specifying the terms for sale and delivery of hydrogen~\cite{aurora2024bankability}. In essence, an HPA functions much like a power purchase agreement in the electricity sector, but for hydrogen fuel. An HPA typically defines several core elements such as price, volume, and delivery conditions. The price term often locks in a fixed price (or price formula) per unit of hydrogen for the duration of the contract. The volume commitment requires the offtaker to purchase a specified annual quantity of hydrogen (e.g., tonnes per year), often under take-or-pay provisions ensuring payment for a minimum volume regardless of actual offtake.

The primary purpose of an HPA is to provide long-term revenue certainty to hydrogen producers, thus making green hydrogen projects \textit{bankable}. By securing a buyer and fixing the hydrogen price, an HPA generates a predictable cash flow that enables lenders to provide debt. The HPA shares risk between the producer and the offtaker; the producer gains revenue assurance with the price risk largely transferred to the buyer, and the offtaker secures a reliable hydrogen supply for its operations or end-users. This structure is especially critical in the green hydrogen sector’s early stages, when high production costs and uncertain demand could otherwise undermine investment.

Furthermore, governments and international agencies often link HPA-backed projects with support schemes; a robust HPA can be a prerequisite for accessing green hydrogen production subsidies or contracts-for-difference that bridge the green premium~\cite{aurora2024bankability}. In summary, by providing predictable revenue over horizons that exceed ten years, HPAs make green hydrogen projects financeable, attract the investment required for final investment decisions, and mitigate the risks that would otherwise deter investment in the nascent hydrogen market. Germany’s H2Global initiative exemplifies how HPAs can unlock green hydrogen investments \cite{h2global2022}.

Hydrogen Purchase Agreements introduce a number of interrelated mathematical and optimisation challenges for developers, particularly when production is driven by variable renewable energy sources such as wind or solar. These challenges are critical to ensuring contract feasibility, pricing adequacy, and operational efficiency.

\begin{itemize}
\item Contract volume determination under uncertainty

The first problem involves determining the committed delivery volume to the offtaker. Since hydrogen production is directly coupled to intermittent electricity generation, the developer must account for production variability when setting the annual contracted quantity. In practice, this leads to a risk-averse optimisation problem, where the goal is to choose a volume that is lower than the expected annual output, ensuring that the committed quantity can be reliably delivered even under unfavourable production scenarios (e.g., low wind years). A simple method for determining the contract volume is presented in Section~\ref{S:Contract volume determination}.

\item Contract price optimisation

The second challenge is determining a reasonable HPA price that balances project bankability and competitiveness. From the developer’s perspective, this involves modelling the levelised cost of hydrogen, accounting for capital and operational expenses, as well as estimating market price evolution for hydrogen. The contract price must provide sufficient return on investment while remaining attractive to the offtaker. Although the contract price is important, this research does not delve into optimisation methods for setting such prices.

\item Real-time dispatch and market arbitrage

Once the HPA is signed, developers face a dynamic operational problem. Suppose a producer has an HPA to deliver 90~tonnes of H$_2$/year at €5/kg. Each hour, the operator must decide whether to allocate the hydrogen produced to fulfil the HPA or to sell it on the spot market if a higher price is available. This creates a multi-period stochastic optimisation problem with constraints on cumulative HPA delivery. The objective is to maximise revenue while ensuring that at least 90~tonnes are delivered by the end of the year. This requires anticipating market prices and managing inventory using optimisation.
\end{itemize}

Although all three challenges are critical for advancing green hydrogen production, this research focuses on the real-time dispatch and market arbitrage problem---a topic that remains underexplored in the literature. In~\cite{bokde2020graphical}, the sequential daily planning is based on a graphical approach that considers the latest electricity prices and CO$_2$ intensities. However, this approach is unable to incorporate complex system dynamics. The sequential daily planning methods in~\cite{FARAH_green_hydrogen, FARAH_green_hydrogen_conference} use a co-optimisation approach that considers a shrinking window of recent electricity prices and CO$_2$ intensities alongside short-term forecasts of those same variables. Although these methods can incorporate complex system dynamics, the methods rely on a single set of input time series, which may cause the optimisation output to be overly sensitive to inputs, adversely affecting cost and CO$_2$ emissions.

To improve sequential daily planning, this research introduces Bounded Fuzzy Logic Control (BFLC) for hydrogen production. Fuzzy logic control is adopted for its simplicity, interpretability, and capacity to accommodate complex systems, as demonstrated in studies in various fields of research, such as wind turbines~\cite{Soliman_Adaptive_Fuzzy_Logic_wind_turbine}, refrigeration systems~\cite{Belman-Flores_Fuzzy_Logic_Control_refrigeration_systems}, steering control~\cite{Arifin_steering_Fuzzy_Logic_Control}, and energy storage control~\cite{Liu_fuzzy_logic_control_hydrogen_production}. The fuzzy logic controller is optimised to determine the daily volume of hydrogen to be exported under the HPA. When necessary, this daily volume is adjusted to ensure that cumulative hydrogen exports to the HPA remain within predefined time-dependent upper and lower bounds. These bounds prevent hydrogen exports to the HPA from progressing either too rapidly or too slowly. The daily volume determined by this process is taken into consideration in a subsequent dispatch optimisation that allocates the hourly energy and hydrogen flows.

In addition to the fundamentally different structure of the BFLC compared to the approaches in~\cite{bokde2020graphical, FARAH_green_hydrogen, FARAH_green_hydrogen_conference}, this research examines a system that exhibits two additional features: the first is the exclusive reliance on wind energy for hydrogen production, with the import of electricity from the grid prohibited. The second is the integration of not only the electricity day-ahead market but also a dynamic hydrogen market in the sequential daily planning and dispatch of energy flows. These features complicate the system control, thereby demonstrating the capability of the BFLC to deliver reliable, market-responsive hydrogen scheduling under stringent, wind-only conditions. 

The subsequent sections of the paper are structured as follows. Section~\ref{S:System description} describes the hydrogen production system and the overall control framework. Section~\ref{S:Contract volume determination} details the calculation of the contract volume, a prerequisite target for controlling the hydrogen plant. Section~\ref{S:Benchmark control} introduces the benchmark control strategy that achieves optimal performance; its outputs are used to train the BFLC. Section~\ref{S:Bounded fuzzy logic control} outlines the operation logic, optimisation, and bounding of the fuzzy logic controller. Section~\ref{S:Results} presents an economic analysis that demonstrates the benefits of the BFLC. Section~\ref{S:Summary and conclusions} summarises the research key findings and conclusions. Finally, the appendices provide supporting data and methodological details.

\section{System description}
\label{S:System description}

\begin{figure}
    \centering
    \includegraphics[width=0.9\columnwidth]{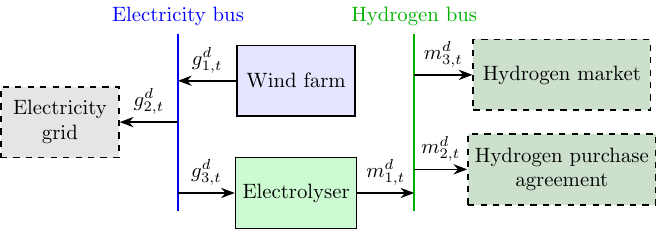}
    \caption{Illustration of electricity and hydrogen flows between the grid, wind farm, electrolyser, and off-takers.}
    \label{fig:Green hydrogen production plant}
\end{figure}

The green hydrogen production plant consists of a 2~MW wind farm, a 1~MW electrolyser, an electricity bus, and a hydrogen bus. These buses are connected to the electricity grid, the HPA stream, and the hydrogen market stream, as illustrated in Figure~\ref{fig:Green hydrogen production plant}. The renewable energy generated by the wind farm ($g_{1,t}^d$) is directed to the electricity bus, which then distributes the energy simultaneously to the electricity grid ($g_{2,t}^d$) and the electrolyser ($g_{3,t}^d$). The connection capacity to the grid is assumed to be sufficiently large to ensure unrestricted electricity flow. Technically, electricity can be imported from the grid for green hydrogen production when electricity prices are sufficiently low. However, the adopted system deliberately excludes grid imports to ensure that the hydrogen produced is exclusively derived from on-site renewable electricity, thereby guaranteeing compliance with the latest European regulations on green hydrogen~\cite{Directorate_General_for_Energy}.  

Wind energy supplied to the electrolyser generates renewable green hydrogen ($m_{1,t}^d$), which is delivered to the hydrogen bus. The specific energy for hydrogen production is 57.6~kWh/kg, consisting of 53.6~kWh/kg for water splitting in the electrolyser and an additional 4~kWh/kg for compression to high pressure~\cite{danish_energy_agency2024, awoe_hydrogen_compression2022}. The hydrogen is distributed in two distinct streams: ($m_{2,t}^d$) directed to the HPA and ($m_{3,t}^d$) allocated to the hydrogen market. Both connections to the HPA and the hydrogen market are assumed to have sufficiently large capacities, ensuring unrestricted hydrogen flow. However, hydrogen flow between the HPA and the hydrogen market is not allowed.

The system is operated with the requirement of supplying a specified quantity of hydrogen $({M}_2^*)$ to the HPA by the end of the year. This quantity should be compatible with the availability of wind resources and the characteristics of the hydrogen production plant. On the one hand, an excessively high ${M}_2^*$ target might be unachievable if wind resources are low, and it reduces flexibility to trade electricity and hydrogen. On the other hand, a low ${M}_2^*$ reduces the risk of not producing the required amount of hydrogen and increases the trading flexibility. However, a low ${M}_2^*$ also reduces the guaranteed revenue from the HPA and increases the risk of not achieving financial benefits if the prices of electricity and hydrogen are low. Furthermore, a low ${M}_2^*$ increases the proportion of uncertain cash flows in the business case, thus reducing the overall bankability. Therefore, a balanced ${M}_2^*$ is required; the calculation of ${M}_2^*$ is presented in Section~\ref{S:Contract volume determination}.

Once ${M}_2^*$ is specified, the challenge is to determine the amount of hydrogen $\overline{}{M}_{2}^d$ that should be exported to the HPA on any specific day \(d\). This challenge is the main focus of this research which presents Bounded Fuzzy Logic Control (BFLC) to determine $\overline{M}_{2}^d$ as detailed in Section~\ref{S:Bounded fuzzy logic control}. The last step in the control of the hydrogen production plant is the dispatch optimisation that determines the energy and hydrogen flows for every hour of day \(d\). The disptach optimisation considers $\bar{M}_{2}^d$ as a hydrogen demand that should be prioritised and maximises the revenue from selling both electricity and hydrogen in their respective markets. The revenues are calculated taking into account the time series of electricity and hydrogen prices.
The dispatch optimisation does not include the revenue from the HPA as this revenue is regarded as a predetermined fixed lump sum for the entire year; hence, this revenue is unaffected by the dispatch optimisation and by the timing of exporting hydrogen to the HPA.

\nomenclature{BFLC}{Bounded fuzzy logic control}

The green hydrogen plant model is developed in Python for Power System Analysis (PyPSA), which is an open-source framework for the simulation and optimisation of power and energy systems~\cite{PyPSA}. The model, which maintains the energy and mass balances for every component of the green hydrogen plant, is optimised by the state-of-the-art mathematical programming solver Gurobi~\cite{gurobi}. The input time series of the model consist of electricity prices, hydrogen prices, and wind capacity factors for years 2015--2023 in Denmark. A sample of these time series is presented in Appendix~\ref{S:Time series of electricity prices, hydrogen prices, and wind capacity factors}. A subset of the times series corresponding to years 2015, 2016, and 2023 is utilised for out-of-sample performance analysis.

\section{Contract volume determination}
\label{S:Contract volume determination}
The contract volume depends on the maximum production of green hydrogen, which is achieved by supplying the electrolyser with as much wind energy as possible, regardless of the prices of electricity and hydrogen. Consequently, the maximum production of green hydrogen (Table~\ref{tab:Maximum hydrogen production}) is limited by the availability of wind resources and the capacities of the wind farm and the electrolyser.
However, the contract volume can only be a fraction of the maximum production of green hydrogen for the financial and risk mitigation considerations discussed in Section~\ref{S:System description}. In this research, 40\% of the mean maximum production of green hydrogen for years 2017--2022 is assumed to be suitable for financial and risk mitigation considerations.
Consequently, the contract volume ${M}_{2}^*$ is 48~tonnes, which represents a substantial amount of hydrogen allocated to the HPA and maintains sufficient operational flexibility to benefit from the trading of electricity and hydrogen in their respective markets. This ${M}_{2}^*$ value is adopted for all years and control approaches, including the benchmark control described in Section~\ref{S:Benchmark control}.

\begin{table}
    \centering
    \caption{Maximum production of green hydrogen}
    \begin{tabular}{cc}
        \hline
        \textbf{Year} & \textbf{H$_2$ (t)} \\
        \hline       
        2017 & 122.7 \\        
        2018 & 117.3 \\        
        2019 & 123.2 \\        
        2020 & 120.6 \\      
        2021 & 115.6 \\       
        2022 & 120.9 \\
        \hline
        mean & 120.0 \\
        ${M}_{2}^*$ & 48.0\\
        \hline
    \end{tabular}
    \label{tab:Maximum hydrogen production}
\end{table}


\section{Benchmark control}
\label{S:Benchmark control}

\begin{figure}
    \centering
    \begin{subfigure}[b]{0.45\textwidth}
        \centering
        \includegraphics[trim=0.5cm 0.0cm 1.5cm 0cm, clip, width=\linewidth]{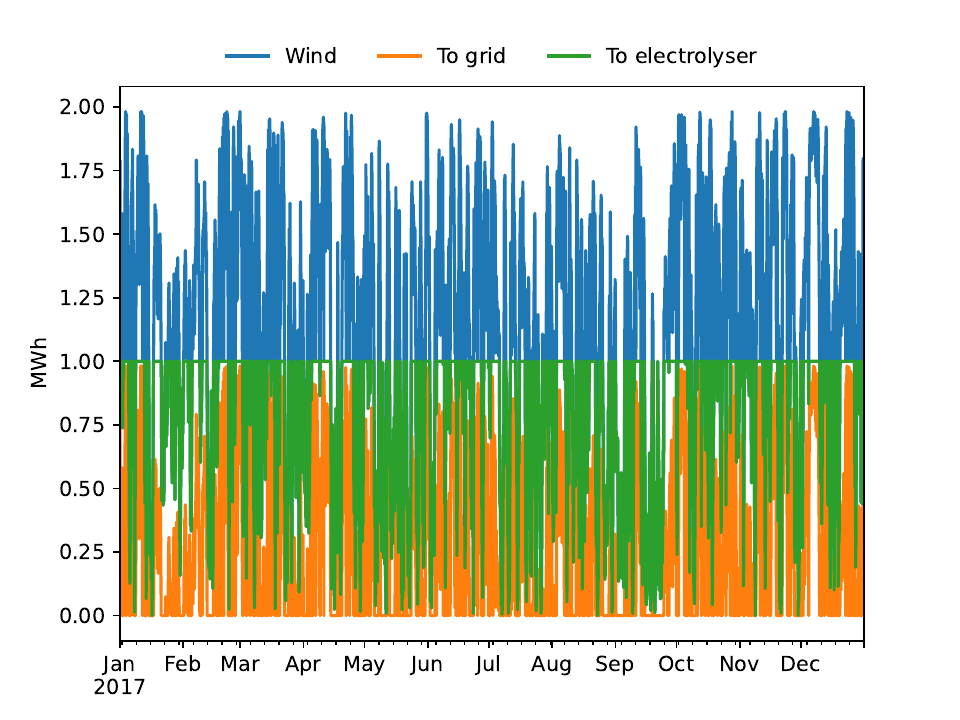}
        \caption{Electrical energy (hourly resolution)}
        \label{fig:Electrical energy (hourly resolution)}
    \end{subfigure}
    \hfill
    \begin{subfigure}[b]{0.45\textwidth}
        \centering
        \includegraphics[trim=0.5cm 0.0cm 1.5cm 0cm, clip, width=\linewidth]{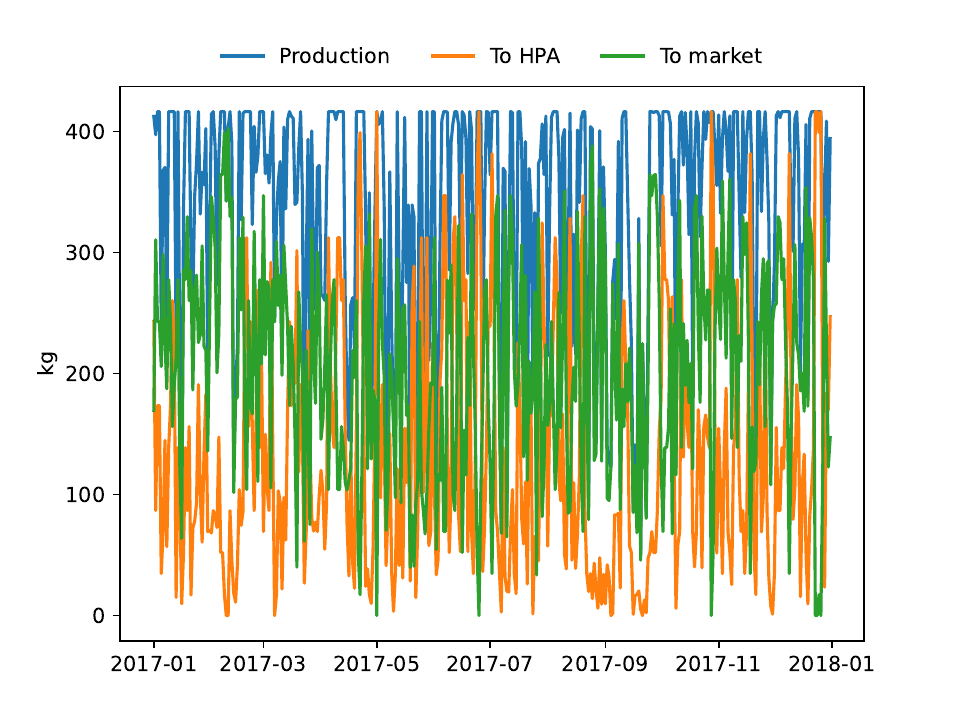}
        \caption{Hydrogen (daily resolution)}
        \label{fig:Hydrogen (daily resolution)}
    \end{subfigure}
    \caption{Sample of benchmark time series based on the perfect foresight of wind capacity factors and prices of electricity and hydrogen for 2017. (a) optimal wind energy production (blue), wind energy export to the grid (orange), and wind energy supply to the electrolyser. (b) Optimal total hydrogen production (blue), hydrogen exports to the HPA (orange), and hydrogen exports to the market (green).}
    \label{fig:Benchmark time series}
\end{figure}

The benchmark performance considers perfect foresight of the input time series, which enables achieving the maximum possible revenue in a given year (Table~\ref{tab:benchmark_revenue}). The benchmark performance optimises the distribution of wind energy between the electricity grid and the electrolyser (Figure~\ref{fig:Electrical energy (hourly resolution)}), as well as the allocation of hydrogen to the HPA and the hydrogen market streams (Figure~\ref{fig:Hydrogen (daily resolution)}). The optimisation is solved with an hourly resolution, however, for effective visualisation, a daily resolution is adopted in Figure~\ref{fig:Hydrogen (daily resolution)}.

\begin{table}
    \centering
    \caption{Benchmark total revenue}
    \begin{tabular}{cc}
        \hline
        \textbf{Year} & \textbf{Total revenue (M€)} \\
        \hline
        2017 & 0.304 \\
        2018 & 0.314 \\
        2019 & 0.327 \\
        2020 & 0.291 \\
        2021 & 0.581 \\
        2022 & 1.493 \\
        \hline
    \end{tabular}
    \label{tab:benchmark_revenue}
\end{table}

The sample benchmark results show that the electrical energy output from the wind farm is capped at its capacity of 2~MW. Similarly, the electrical energy supplied to the electrolyser is limited by its capacity of 1~MW. The electrical energy supplied to the grid does not exceed 1~MW, even though the grid connection is uncapped. This result indicates that electrical energy is prioritised for the electrolyser over the grid, except in instances of excessively high electricity prices. The sample benchmark results also show that hydrogen production is limited by the electrolyser capacity and that hydrogen is supplied to both the HPA and the market almost daily, with only a few exceptions when hydrogen is supplied to either the HPA or the market. The benchmark time series for years 2017--2022 are utilised to train the BFLC as described in Section \ref{S:Bounded fuzzy logic control}.

\section{Bounded fuzzy logic control}
\label{S:Bounded fuzzy logic control}

\subsection{Description}

\begin{figure}
    \centering
    \includegraphics[width=0.9\columnwidth]{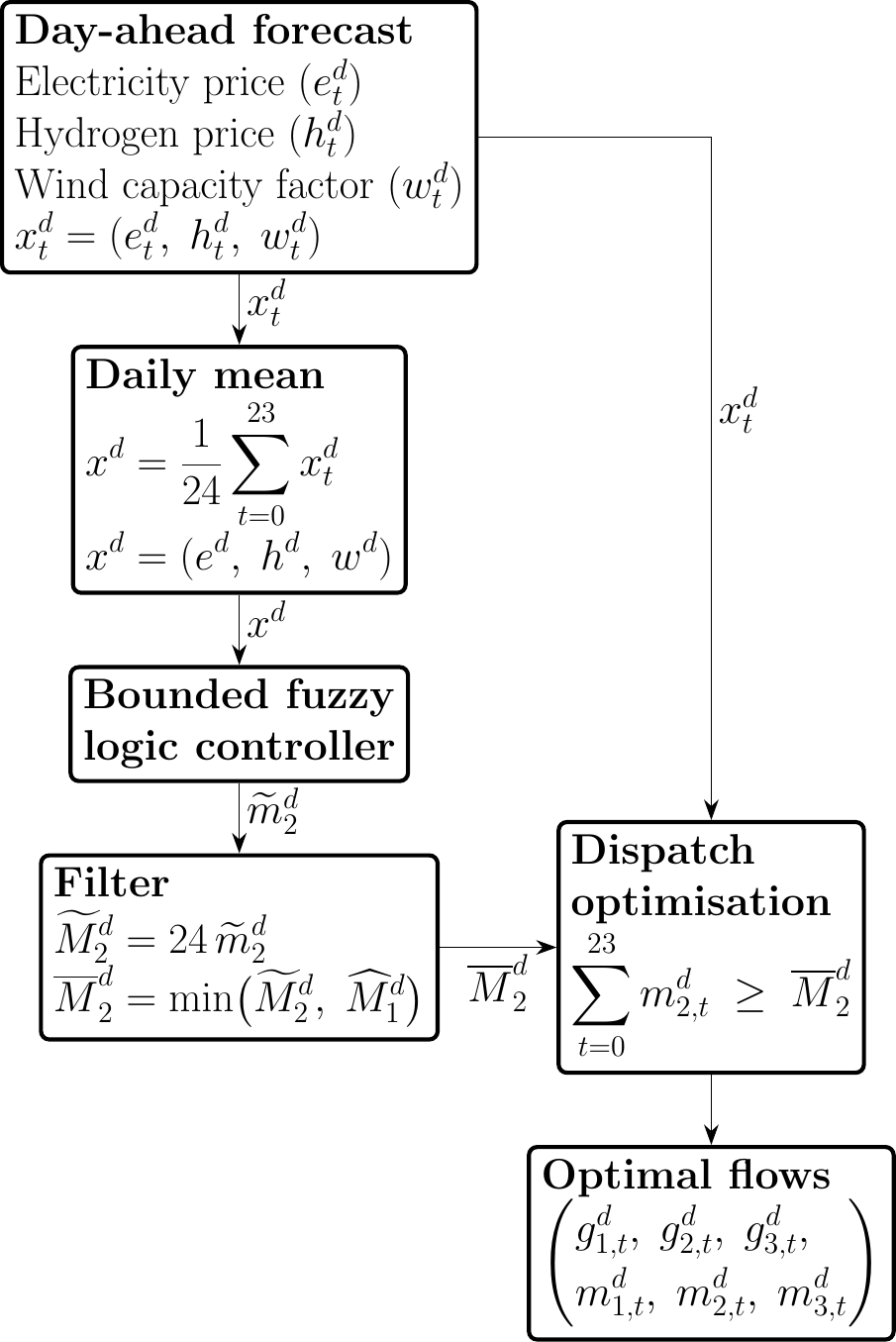}
    \caption{Plant control logic. The hourly day-ahead forecasts are converted to daily mean values that allow the BFLC to determine the daily target of hydrogen export to the HPA ($\widetilde m_2^d$). This target is capped by the maximum possible hydrogen production depending on wind availability. The dispatch optimisation considers the capped target ($\overline M_2^d$) as well as the hourly day-ahead forecast to determine the optimal flows of electricity and hydrogen that maximise revenue. This process is repeated daily.}
    \label{fig:control_logic}
\end{figure}

Unlike the benchmark operation, which is based on the unrealistic perfect foresight of inputs, the operation of the system with fuzzy logic control provides realistic system performance based on sequential decision making where only 24~h of forecast inputs are available (Figure~\ref{fig:control_logic}). The inputs of the fuzzy logic controller are the daily mean electricity price $(e^d)$, the daily mean hydrogen price $(h^d)$, and the daily mean wind capacity factor $(w^d)$. The output of the fuzzy logic controller is the daily mean ($\widetilde{m}_2^d$) of the hydrogen supply target to the HPA ($\widetilde{M}_2^d$). The target $\widetilde{M}_2^d$ is limited to the maximum possible production of green hydrogen on day \(d\) ($\widehat{M}_1^d$) to ensure that the system is capable of producing the required amount of hydrogen $\overline{M}_2^d$. The amount of hydrogen $\overline{M}_2^d$ is set as a minimum limit for the operation of the system on day \(d\), and a typical dispatch optimisation determines the optimal electrical energy and hydrogen flows for every hour of day \(d\). The dispatch optimisation is similar to that of the benchmark operation, except that this optimisation is performed daily with a constraint on $\overline{M}_2^d$ whereas the benchmark operation is performed for a whole year at once with the constraint of supplying $M_2^*$ to the HPA.


\nomenclature{$x_t^d$}{Vector of day-ahead forecasts for hour $t$ and day $d$}
\nomenclature{$x^d$}{Vector of mean day-ahead forecasts for day $d$}

\begin{figure}
    \centering
    \includegraphics[width=\linewidth]{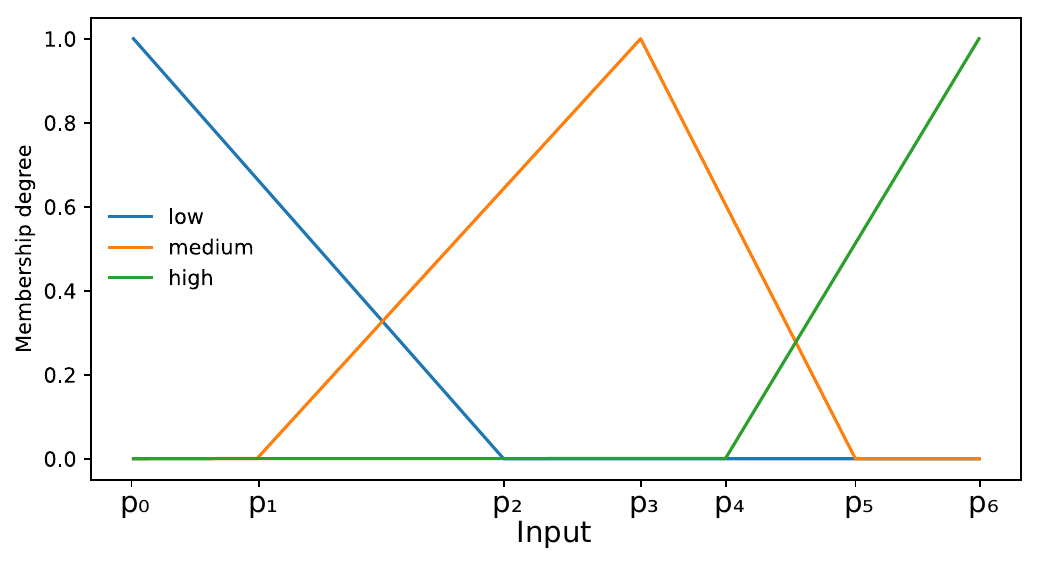}
    \caption{Parametrised triangular membership functions for the fuzzy controller. The "low" function (blue) is defined by the parameters $p_0$ and $p_2$, the "medium" function (orange) is defined by $p_1$, $p_3$, and $p_5$, and the "high" function (green) is defined by $p_4$ and $p_6$. Optimal parameter values for each variable are presented in Table~\ref{tab:parameters}.}
    \label{fig:Parametrised membership functions}
\end{figure}

The fuzzy logic control is implemented using the Python scikit-fuzzy library~\cite{warner2024scikitfuzzy}. Each of the three inputs and the output variables are classified as "low", "medium", and "high" fuzzy sets that are defined by the parametrised triangular membership functions (Figure~\ref{fig:Parametrised membership functions}). The parameters ($p_0$, $p_2$), ($p_1$, $p_3$, $p_5$), and ($p_4$, $p_6$) define the membership functions of the fuzzy sets "low", "medium", and "high", respectively. The parameters $p_0$ and $p_6$ are the minimum and maximum variable values, respectively. The parameters $(p_1, ..., p_5)$ and the fuzzy rules are optimised as presented in Section~\ref{S:Optimisation}.

The fuzzy rules are "if-then" statements that connect the inputs with the "and" operator to determine a fuzzy set for the output. For instance, a fuzzy rule may be as follows:

\begin{quote}
    \texttt{if $e^d$ is low, and $h^d$ is medium, and $w^d$ is high, then $\widetilde{m}_2^d$ is low}
\end{quote}

This rule essentially states that if the electricity price is low, the hydrogen price is medium, and the available wind energy is high, then the hydrogen export to the HPA should be low. The calculation of a numerical output from the fuzzy rules follows the typical centroid method for defuzzification~\cite{ZADEH1965338, ross2005fuzzy}. The total number of possible rules is 81, which corresponds to all combinations of fuzzy inputs and outputs. However, this set of rules includes conflicting entries; a single combination of fuzzy inputs may be linked to different fuzzy outputs. Consequently, optimising the fuzzy rules entails identifying an appropriate rule base that encompasses all combinations of fuzzy input variables, with each combination being associated with a single fuzzy output, either “low”, “medium”, or “high”.

\subsection{Optimisation}
\label{S:Optimisation}

Considering that the results from the benchmark optimisation represent the best distributions of electrical energy and hydrogen flows, these results can be utilised to optimise the fuzzy logic controller. The optimisation process is implemented in Python using the pyswarm package, which enables the application of particle swarm optimisation (PSO)~\cite{pyswarm}. Particle swarm optimisation is a population-based metaheuristic algorithm inspired by the collective behaviour of bird flocks and fish schools. The PSO is well-suited for nonlinear optimisation problems, which makes it an appropriate choice for fuzzy logic control optimisation. The following steps outline the procedure for the fuzzy logic control optimisation. This optimisation considers multiple years of training data to determine the optimal membership functions and fuzzy rules. 

\begin{enumerate}
    \item The PSO algorithm initialises/updates the parameters $(p_1, \dots, p_5)$, thereby defining the fuzzy membership functions for both the inputs and the output.
    
    \item Calculate the fuzzy membership values for the inputs and output based on the numerical inputs and the benchmark time series.
    
    \item Compute the activation degree for each rule for every entry in the time series, following the method described in~\cite{Generating_fuzzy_rules_by_learning_from_examples}. 
    
    \item Sum the activation degrees for each rule.
    
    \item Within each group of conflicting rules, maintain the rule with the highest cumulative degree. This procedure reduces the number of rules from 81 to 27, ensuring that the selected rules are the most consistent with the benchmark results.
    
    \item For each day, compute the output of the fuzzy logic controller using the current membership functions and the rule base. The result is a time series of $\widetilde{m}_2^d$ values.
    
    \item Calculate the optimisation objective function as the sum of two components: the first is the sum of squared difference between the benchmark hydrogen supply to the HPA and the output ($\widetilde{m}_2^d$) of the fuzzy logic controller. The second is the squared difference between the sum of benchmark hydrogen supply to the HPA and the sum of $\widetilde{m}_2^d$.
    
    \item If the PSO stopping criteria are unmet; neither an optimal solution is identified nor the maximum number of iterations (100 iterations) is reached, restart from step 1. If a stopping criterion is met, repeat steps 2--6 using the parameters calculated by the PSO.
\end{enumerate}

\begin{figure*}
    \centering
    \begin{subfigure}[b]{0.45\textwidth}
        \includegraphics[trim=1cm 0.5cm 1cm 1.2cm, clip, width=\linewidth]{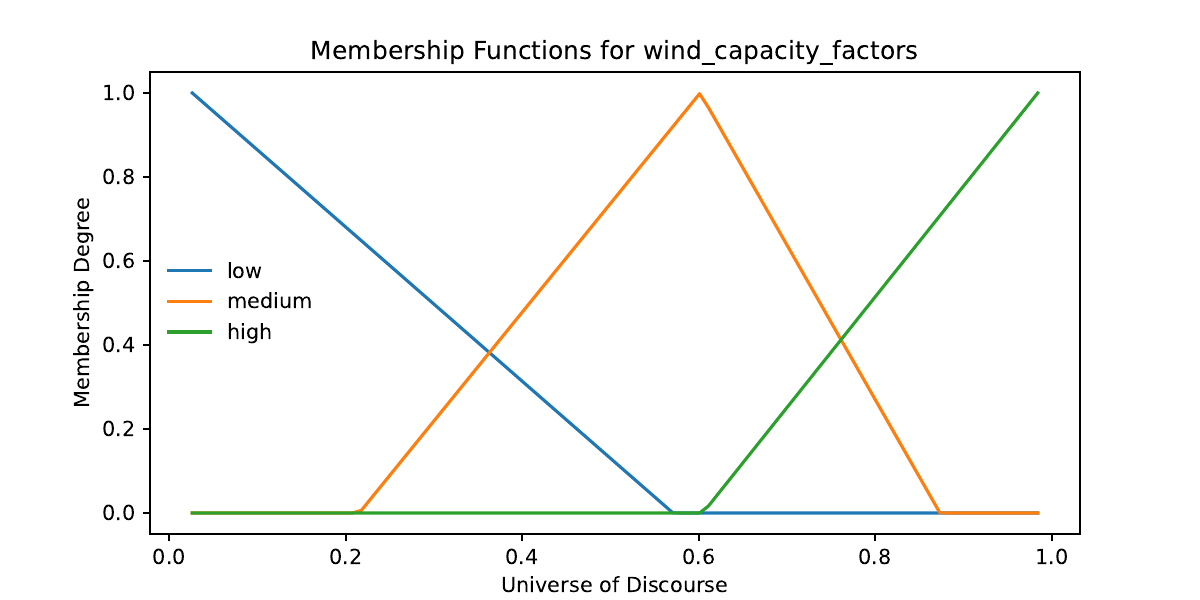}
        \caption{Wind capacity factor (input)}
        \label{fig:Membership function of wind capacity factor_h2_to_hpa}
    \end{subfigure}
    \hfill
    \begin{subfigure}[b]{0.45\textwidth}
        \includegraphics[trim=1cm 0.5cm 1cm 1.2cm, clip, width=\linewidth]{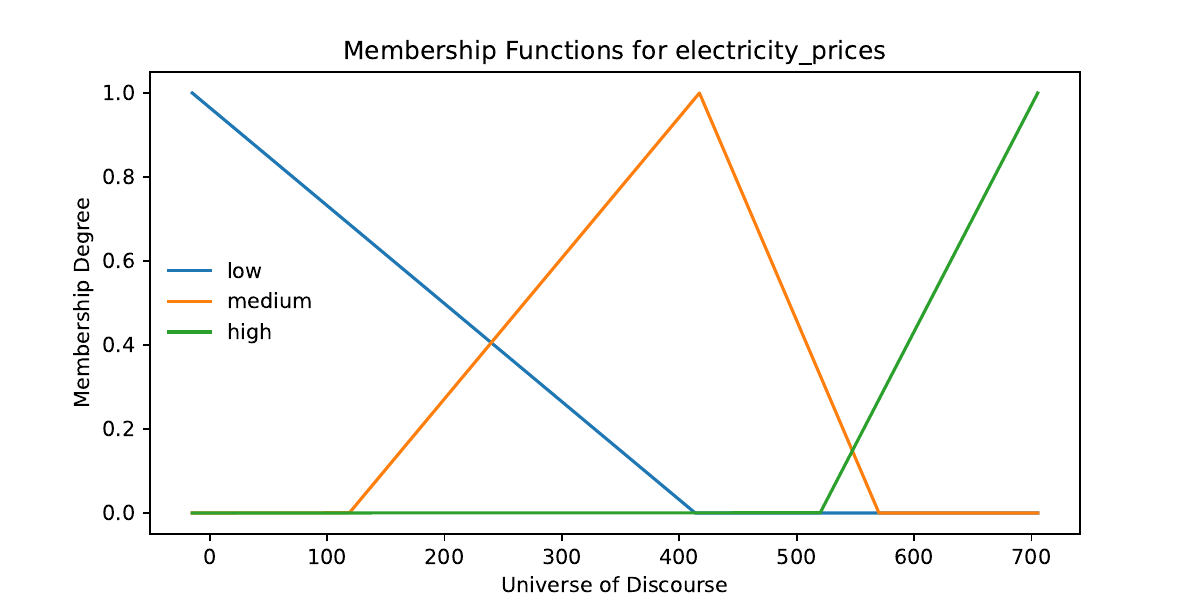}
        \caption{Electricity price (input)}
        \label{fig:Membership function of electricity price_h2_to_hpa}
    \end{subfigure}
    \hfill
    \begin{subfigure}[b]{0.45\textwidth}
        \includegraphics[trim=1cm 0.5cm 1cm 1.2cm, clip, width=\linewidth]{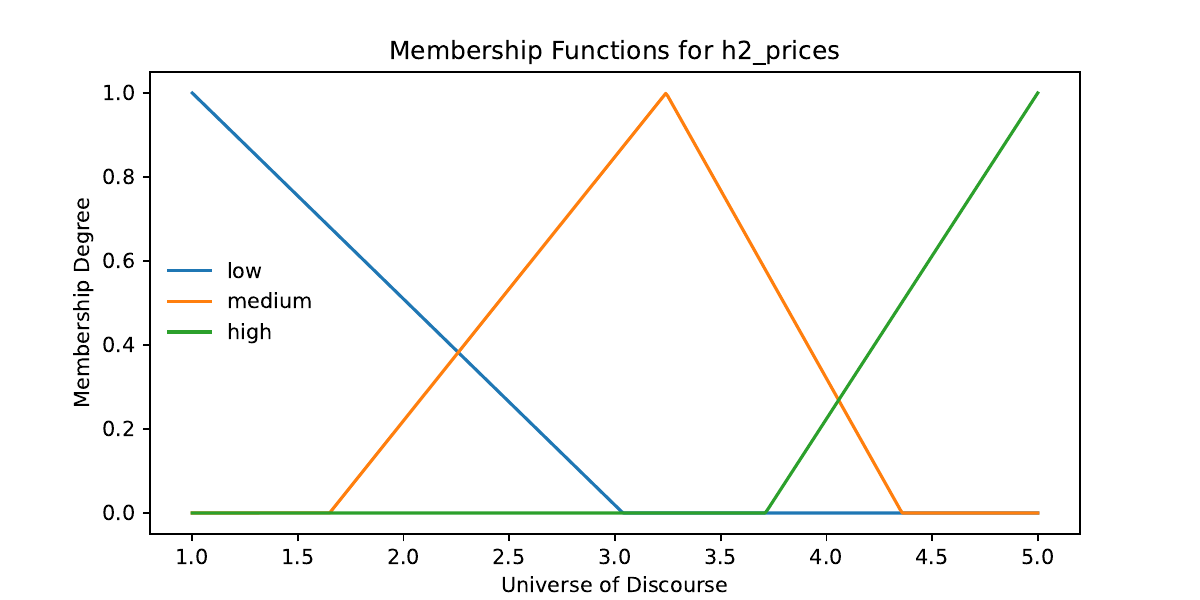}
        \caption{Hydrogen price (input)}
        \label{fig:Membership function of hydrogen price_h2_to_hpa}
    \end{subfigure}
    \hfill
    \begin{subfigure}[b]{0.45\textwidth}
        \includegraphics[trim=1cm 0.5cm 1cm 1.2cm, clip, width=\linewidth]{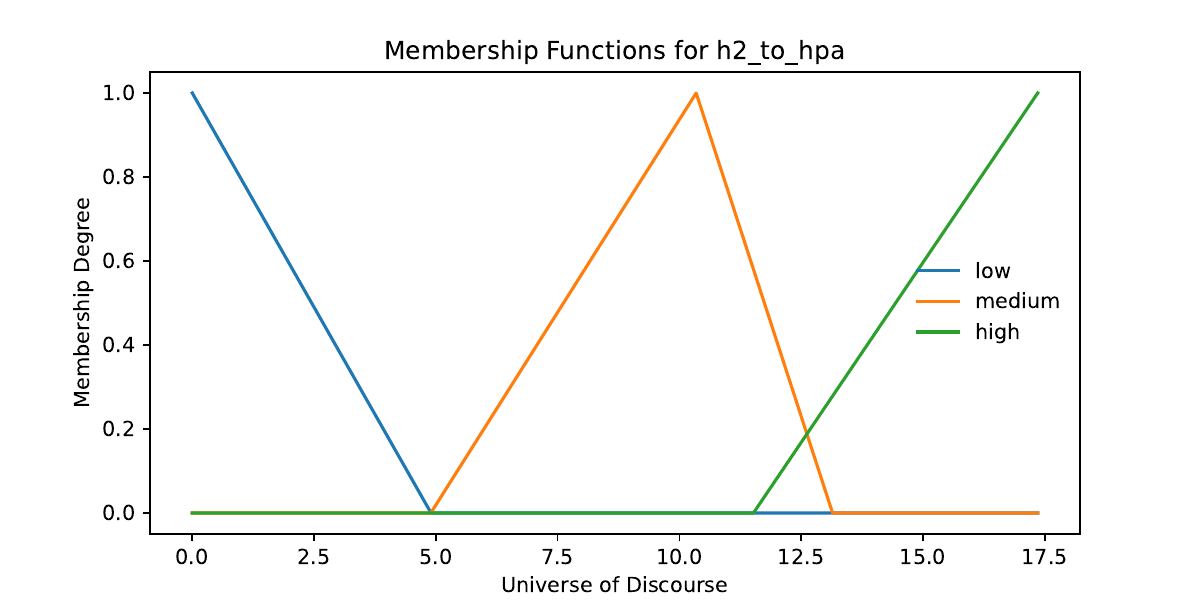}
        \caption{Hydrogen supply to hpa (output)}
        \label{fig:Membership function of hydrogen supply to HPA}
    \end{subfigure}
    \caption{Membership functions of the fuzzy inputs and output of the controller of hydrogen supply to the HPA}
    \label{fig:Membership functions hpa}
\end{figure*}

This optimisation provides the optimal membership functions and the rule base. The optimal parameters in Table~\ref{tab:parameters} define the membership functions in Figure~\ref{fig:Membership functions hpa}, and the most suitable rules defining the rule base are summarised in Table~\ref{tab:fuzzy_rules}.

\begin{table}
    \centering
    \caption{Optimal parameters of membership functions}
    \label{tab:parameters}
    \begin{tabular}{lccccccc}
        \toprule
        \textbf{} & \textbf{p$_0$} & \textbf{p$_1$} & \textbf{p$_2$} & \textbf{p$_3$} & \textbf{p$_4$} & \textbf{p$_5$} & \textbf{p$_6$}\\
        \midrule
        w$^d$ & 0.01 & 0.22 & 0.57 & 0.60 & 0.60 & 0.87 & 0.99\\
        e$^d$ & -11.76 & 118.85 & 413.26 & 417.15 & 519.96 & 569.89 & 695.09\\
        h$^d$ & 1.04 & 1.65 & 3.04 & 3.24 & 3.71 & 4.36 & 5.0\\
        $\widetilde{m}_2^d$ & 0.0 & 4.90 & 4.90 & 10.34 & 11.52 & 13.14 & 17.36\\
        \bottomrule
    \end{tabular}
\end{table}





\begin{table}
\centering
\caption{Fuzzy rules}
\begin{tabular}{lcccc}
\toprule
\multirow{2}{*}{Rule} & \multicolumn{3}{c}{Inputs} & \multicolumn{1}{c}{Output} \\
\cmidrule(lr){2-4}
\cmidrule(lr){5-5}
 & \textbf{$e^d$} & \textbf{$h^d$} & \textbf{$w^d$} & \textbf{$\widetilde{m}_2^d$} \\
\midrule
1 & low & low & low & low \\
2 & low & medium & low & low \\
3 & low & high & low & low \\
4 & medium & low & low & low \\
5 & medium & medium & low & low \\
6 & medium & high & low & low \\
7 & high & low & low & low \\
8 & high & medium & low & low \\
9 & high & high & low & low \\
10 & low & low & medium & high \\
11 & low & medium & medium & low \\
12 & low & high & medium & low \\
13 & medium & low & medium & low \\
14 & medium & medium & medium & low \\
15 & medium & high & medium & low \\
16 & high & low & medium & low \\
17 & high & medium & medium & low \\
18 & high & high & medium & low \\
19 & low & low & high & high \\
20 & low & medium & high & low \\
21 & low & high & high & low \\
22 & medium & low & high & low \\
23 & medium & medium & high & low \\
24 & medium & high & high & low \\
25 & high & low & high & low \\
26 & high & medium & high & low \\
27 & high & high & high & low \\
\bottomrule
\end{tabular}
\label{tab:fuzzy_rules}
\end{table}

\nomenclature{$e^d$}{Mean electricity price on day {$d$} (€/MWh)}
\nomenclature{$h^d$}{Mean hydrogen price on day {$d$} (€/kg)}
\nomenclature{$w^d$}{Mean wind capacity factor on day {$d$} (-)}

\subsection{Bounding}
\label{S:Bounding}

\begin{figure}
    \centering
    \includegraphics[width=\linewidth]{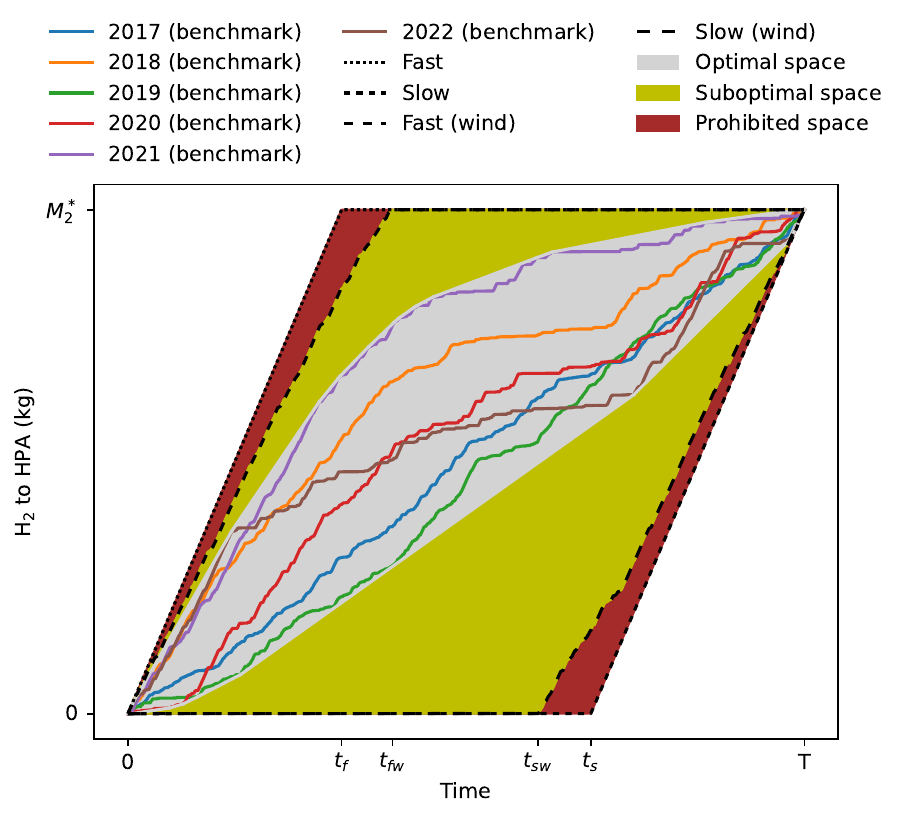}
    \caption{Boundaries of cumulative hydrogen export to the HPA. The optimal space (grey area) is the convex hull of the benchmark cumulative hydrogen export to the HPA for different years. The upper and lower boundaries of the convex hull are the limits of cumulative hydrogen export to the HPA when the fuzzy control is utilised. Without these boundaries, the fuzzy logic control may drive the cumulative hydrogen export to the HPA too quickly or too slowly, crossing from the optimal space to the suboptimal space (yellow area). In this case, the ${M}_2^*$ target might still be reached relying on renewable energy only. If the cumulative hydrogen export reaches the prohibited space (brown area, right side), additional electricity from the grid is needed to reach the ${M}_2^*$ target. The prohibited space on the left side is unreachable without electricity import from the grid. The earliest possible times to reach the ${M}_2^*$ target are $t_f$ and $t_{fw}$ following the "Fast" and "Fast (wind)" paths respectively. In contrast, the latest times to start exporting hydrogen to the HPA and still reach the ${M}_2^*$ target are $t_sw$ and $t_{s}$ following the "Slow (wind)" and "Slow" paths respectively.}
    \label{fig:Boundaries of cumulative hydrogen export to the HPA}
\end{figure}

The optimised fuzzy logic controller is able to allocate suitable daily hydrogen export to the HPA. However, depending on the input conditions, the fuzzy logic controller may consistently export hydrogen to the HPA---risking over-exporting and overlooking better future opportunities---or curtail export, potentially failing to meet the HPA target. In other words, the optimised fuzzy logic controller does not have the sense of being too fast or too slow in exporting hydrogen to the HPA. Therefore, further control of the hydrogen export to the HPA is required to ensure that the cumulative hydrogen export to the HPA is maintained within acceptable boundaries depending on the time of year. Such boundaries can be established from the cumulative hydrogen exported to the HPA based on the benchmark results (Figure~\ref{fig:Boundaries of cumulative hydrogen export to the HPA}).

Exporting the required amount of hydrogen to the HPA can follow an infinite number of paths. For instance, the "Fast" path aims to fulfill the HPA hydrogen target at the earliest possible time by continuously operating the electrolyser at full capacity. In principle, this path is prohibited as the intermittent availability of wind energy requires importing electricity from the grid to maintain the continuous operation of the electrolyser. To fulfill the HPA hydrogen target as quickly as possible relying only on wind energy, the cumulative hydrogen export to the HPA can follow the path "Fast (wind)". This path allocates as much wind energy as possible to the electrolyser to reach the target in the earliest possible time ($t_{fw}$). The path "Fast (wind)" changes slightly from year to year depending on the availability of wind energy. In contrast, the extreme paths "Slow" and "Slow (wind)" can be followed, which postpone hydrogen exports to the HPA for as long as possible.

However, these extreme paths are unsuitable for achieving acceptable performance; the electricity and hydrogen prices, as well as the wind availability, are unlikely to be arranged in a way that favours such extreme paths. This is evident from the benchmark results, which show that the cumulative hydrogen exports to the HPA seem to form a cluster between the extreme paths. Despite the highly irregular patterns, optimal cumulative hydrogen exports to the HPA for different years can be contained within an optimal space formed by the convex hull of the benchmark cumulative hydrogen export to the HPA (Figure~\ref{fig:Boundaries of cumulative hydrogen export to the HPA}). Although the optimal cumulative hydrogen export to the HPA may cross to the outside of the optimal space for some unusual years, the convex hull is likely to encompass most optimal paths. The boundaries of the convex hull can be refined if additional benchmark results are included; nevertheless, with benchmark results from only six years, the convex hull seems to have formed a substantially large core.
In summary, the output of the fuzzy logic controller is unaltered as long as the cumulative hydrogen export to the HPA remains within the boundaries of the optimal space defined by the convex full. If the output of the fuzzy logic controller drives the cumulative hydrogen export to the outside of the optimal space boundaries, the output is modified to either increase or decrease the hydrogen export to the HPA, depending on whether the upper or lower bound is being crossed.



\section{Revenue analysis}
\label{S:Results}

The analysis considers the total revenue from the exports of electricity and hydrogen to their respective markets. In practice, the total revenue also includes the revenue from the hydrogen exported to the HPA, i.e., the contract value. However, as the contract value is a predetermined fixed amount agreed upon by the hydrogen producer and the off-taker, regardless of the control of the system, and as the analysis focuses on revenue comparisons, the contract value is excluded from the calculation. The analysis also considers the normalised total revenues calculated as the ratio of total revenues with steady control and BFLC to the benchmark total revenue. The steady control is based on a constant daily export of hydrogen to the HPA (see Appendix~\ref{S:Steady controller}). This control is utilised for comparison to highlight the benefits of utilising the BFLC.

\begin{figure}
    \centering
    \includegraphics[trim=0.5cm 0.0cm 0.0cm 0.5cm, clip, width=\linewidth]{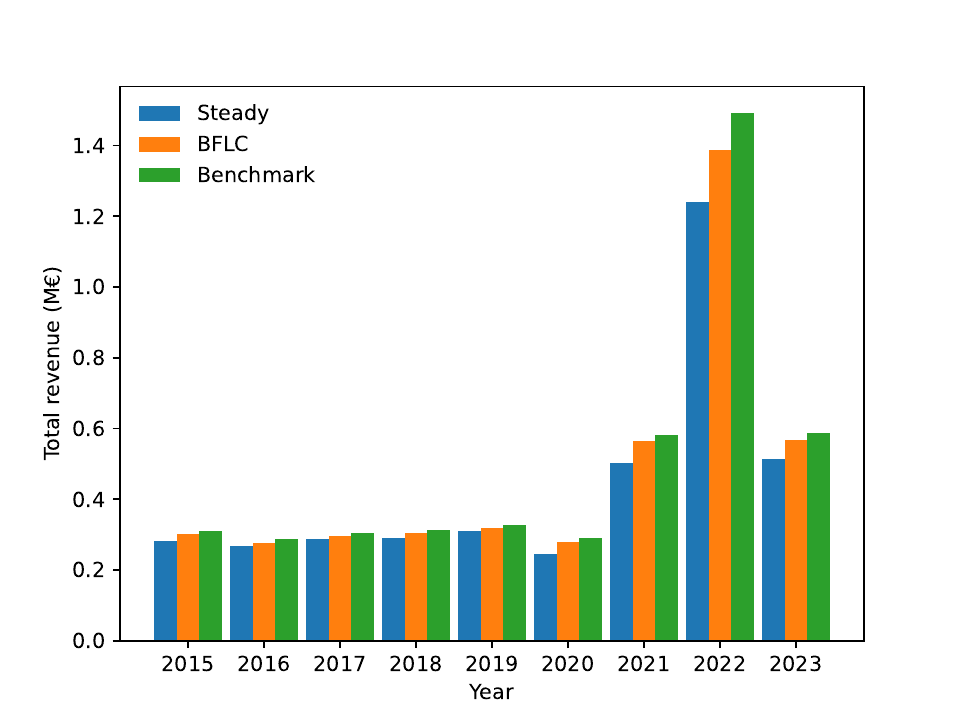}
    \caption{Comparison of total revenues achieved with the steady control (blue), BFLC (orange), and the benchmark approach (green)}
    \label{fig:Total revenue}
\end{figure}

The benchmark total revenue is relatively stable, slightly exceeding 0.3~M€ from 2015 to 2020. However, a significant increase is observed from 2021 to 2023, culminating in a notable peak of 1.5~M€ in 2022 (Figure~\ref{fig:Total revenue}). This substantial increase can be attributed to the increase in electricity prices, which more than doubled during the 2021--2023 period, as detailed in Table~\ref{tab:Electricity prices statistics}. The total revenues achieved by steady control and BFLC follow a pattern similar to that of the benchmark for the same period, with BFLC consistently achieving higher revenues than steady control.

\begin{table}
\centering
\caption{Electricity prices statistics}
\label{tab:Electricity prices statistics}
\begin{tabular}{lcc}
\toprule
\multirow{2}{*}{Year} & \multicolumn{1}{c}{Mean} & \multicolumn{1}{c}{Standard deviation} \\
 & \multicolumn{1}{c}{(€/MWh)} & \multicolumn{1}{c}{(€/MWh)} \\
\midrule
2015 & \phantom{0}22.9 & \phantom{0}11.1 \\
2016 & \phantom{0}26.7 & \phantom{0}\phantom{0}9.8 \\
2017 & \phantom{0}30.1 & \phantom{0}10.7 \\
2018 & \phantom{0}44.1 & \phantom{0}15.1 \\
2019 & \phantom{0}38.5 & \phantom{0}13.2 \\
2020 & \phantom{0}25.0 & \phantom{0}17.4 \\
2021 & \phantom{0}88.1 & \phantom{0}64.8 \\
2022 & 219.0 & 145.5 \\
2023 & \phantom{0}86.8 & \phantom{0}48.8 \\
\bottomrule
\end{tabular}
\end{table}

\begin{figure}
    \centering
    \includegraphics[trim=0.5cm 0.0cm 0.0cm 0.0cm, clip, width=\linewidth]{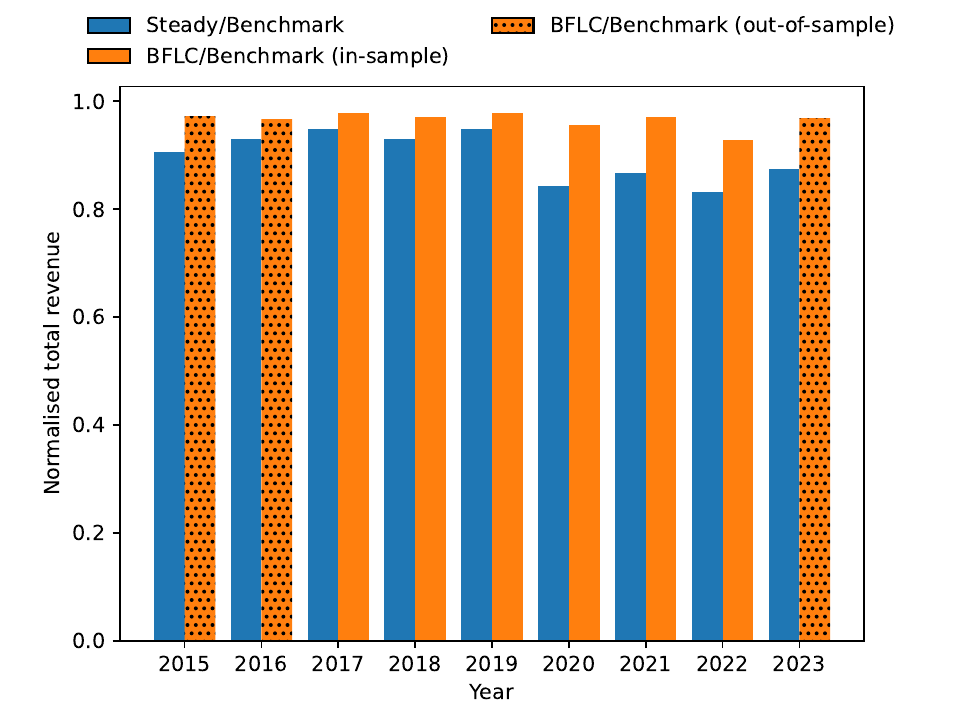}
    \caption{Comparison of normalised total revenues achieved with steady control (blue), in-sample BFLC (orange), and out-of-sample BFLC (orange, black dotted)}
    \label{fig:Normalised total revenue}
\end{figure}

The normalised total revenues achieved with the BFLC are consistently higher than 92\%; the lowest normalised total revenue is 92.8\% achieved in 2022 (Figure~\ref{fig:Normalised total revenue}). This year is characterised by elevated electricity price levels and high variability (Table~\ref{tab:Electricity prices statistics}). The normalised total revenues achieved with steady control are consistently lower than those of the BFLC. A smaller difference is observed in years of low electricity price levels and low variability (2015--2019). Low electricity price levels and low variability reduce the adverse impact of suboptimal control on revenue. However, during periods of high electricity price levels and high variability (2020--2023), the normalised total revenue with steady control is more than 12\% lower than those of the benchmark; the lowest normalised total revenue is 16.8\% lower than that of the benchmark in 2022. To verify that these observations are not unique to the Danish case study, a similar analysis for a site in the United Kingdom shows congruent results (see Appendix~\ref{S:Revenue analysis for a United Kingdom case study}). Collectively, these results demonstrate the advantage of BFLC over steady control, especially under conditions of high electricity price levels and high variability.

The results indicate that benchmark performance--based on perfect foresight--can significantly overestimate the economic benefits of renewable hydrogen production. Such overestimations can lead to unrealistic expectations and increase the risk of financial losses for project developers. In contrast, simpler operational strategies, such as steady control, tend to underestimate potential economic gains. This conservative approach may obscure the actual feasibility of renewable hydrogen projects, potentially discouraging investment. Therefore, adopting an advanced control strategy, such as the BFLC, is essential for a realistic evaluation that mitigates economic risks without overlooking economically viable projects.

\section{Summary and conclusions}
\label{S:Summary and conclusions}

Hydrogen produced from renewable energy sources has the potential to play an important role in the green transition by decarbonising heavy industries. However, green hydrogen projects are facing development challenges related to high costs and underdeveloped demand-side markets. These challenges increase the financial risks for developers of green hydrogen projects and reduce investments in this important sector. To mitigate financial risks and improve revenue predictability, HPAs can be adopted to provide long-term revenue certainty to hydrogen producers. However, the long-term hydrogen production targets specified in HPAs complicate the control of the production plant. The control needs to maximise short-term revenues from selling electricity and hydrogen in their respective markets, while also ensuring that the HPA target can be met at the end of the year. Consequently, this research presented a Bounded Fuzzy Logic Control (BFLC) that can sequentially specify the optimal daily hydrogen quantity that should be dedicated to the HPA target. This quantity was subsequently considered in dispatch optimisation to determine the energy and hydrogen flows for each hour of the day. The hydrogen production plant considered in this research consisted of an electrolyser that was supplied by wind energy. The plant was connected to the electricity grid, hydrogen market stream, and HPA stream.
The BFLC utilised the input variables, namely wind capacity factors, electricity prices, and hydrogen prices to determine the daily hydrogen quantities dedicated to the HPA. The membership functions and fuzzy rules of the BFLC were optimised based on optimal benchmark results that considered perfect foresight of the input variables for the entire year. The benchmark results were also utilised to determine bounds that prevent hydrogen exports to the HPA from progressing too quickly or too slowly.
The total revenue achieved when the hydrogen plan is controlled by the BFLC is within 9\% of that of the benchmark. The BFLC revenue was consistently higher than that of a steady control, with larger revenue differences observed during periods of high electricity levels and variability. These results demonstrate the importance of the BFLC as a tool that can assist developers of green hydrogen projects in quantifying expected revenues based on a realistic advanced control strategy.

\begin{acknowledgments}
This research was partially funded by Ørsted A/S and The Danish Energy Technology Development and Demonstration Program under grant number 64020-2120.
\end{acknowledgments}

\appendix

\section{Time series of electricity prices, hydrogen prices, and wind capacity factors}
\label{S:Time series of electricity prices, hydrogen prices, and wind capacity factors}

\begin{figure}
    \centering
    \begin{subfigure}[b]{0.45\textwidth}
        \centering
        \includegraphics[trim=0.5cm 0.0cm 1.5cm 1.4cm, clip, width=\linewidth]{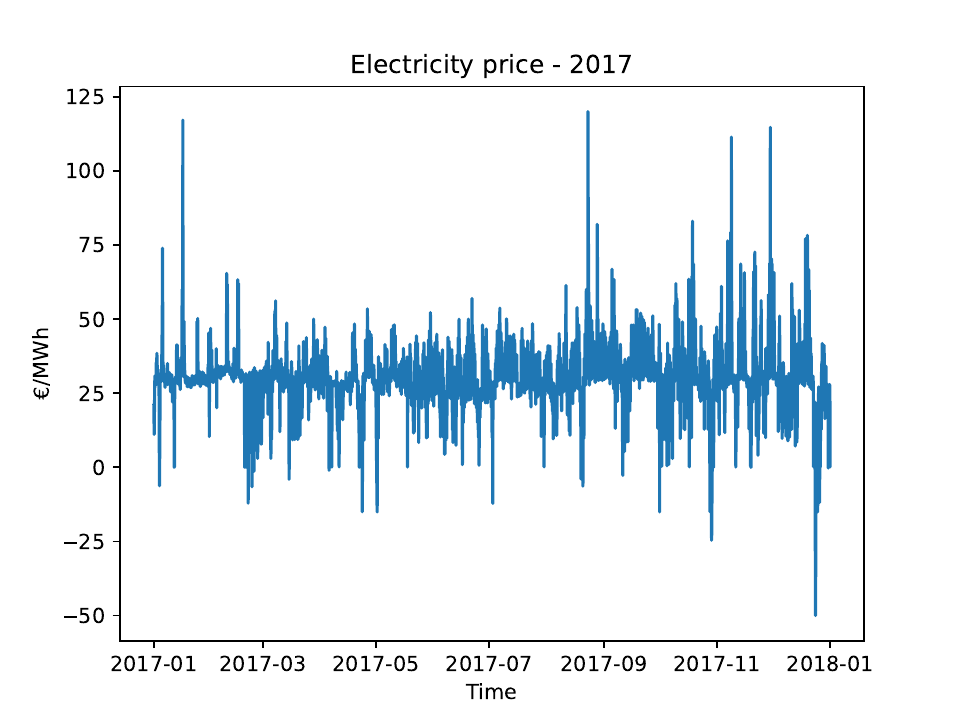}
        \caption{Electricity price, bidding zone: DK1~\cite{entsoe}}
        \label{fig:Electricity price in 2017}
    \end{subfigure}
    \hfill
    \begin{subfigure}[b]{0.45\textwidth}
        \centering
        \includegraphics[trim=0.5cm 0.0cm 1.5cm 1.4cm, clip, width=\linewidth]{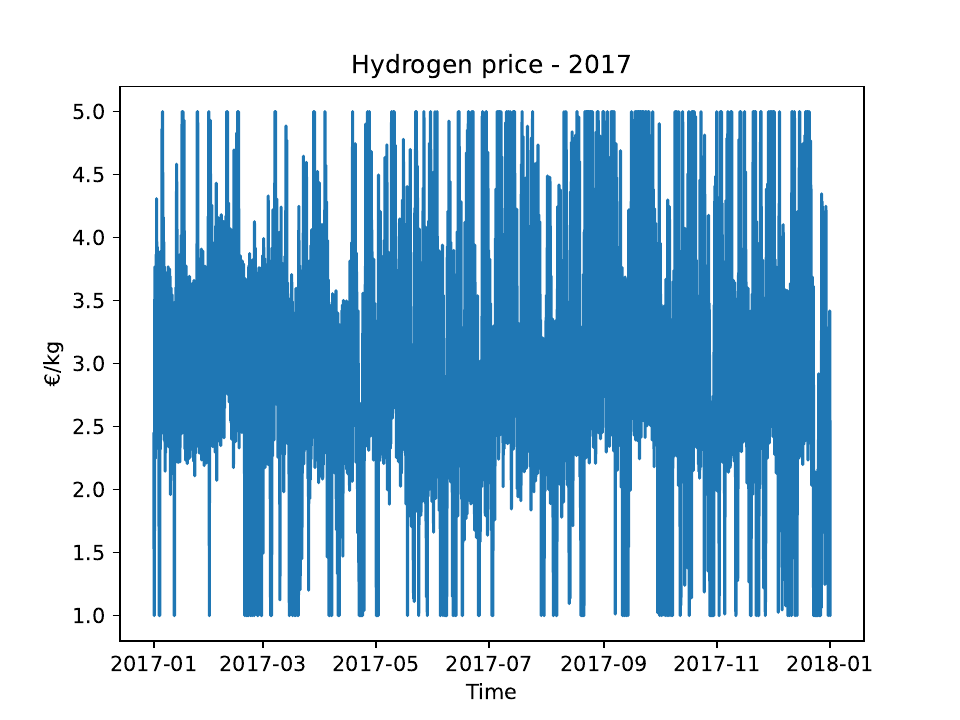}
        \caption{Hydrogen price }
        \label{fig:Hydrogen price in 2017}
    \end{subfigure}
    \hfill
    \begin{subfigure}[b]{0.45\textwidth}
        \centering
        \includegraphics[trim=0.5cm 0.0cm 1.5cm 1.4cm, clip, width=\linewidth]{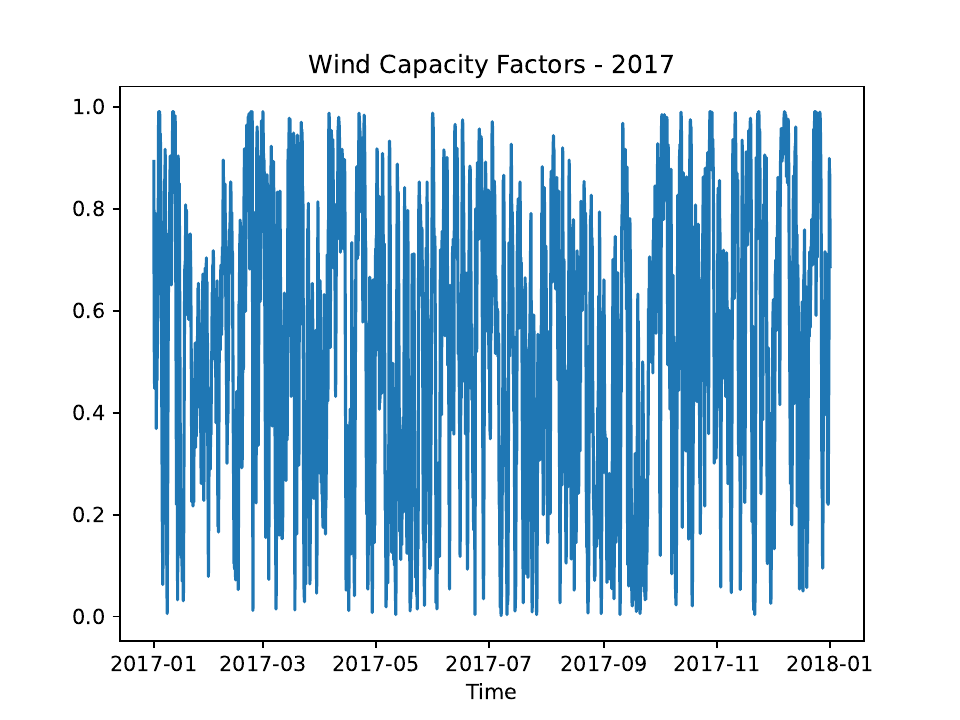}
        \caption{Wind capacity factor, north of Jutland, Denmark, latitude: 56.6135\degree, longitude: 8.9328\degree, dataset: MERRA-2 (global), turbine model: Vestas V90 2000, hub height: 80~m~\cite{RenewablesNinja}}
        \label{fig:Wind capacity factor in 2017}
    \end{subfigure}
    \caption{Input time series (year 2017)}
    \label{fig:prices_and_capacity_factors_2017}
\end{figure}

A sample of the input time series of electricity prices, hydrogen prices, and wind capacity factors is presented in Figure~\ref{fig:prices_and_capacity_factors_2017}. The electricity prices are obtained from the European Network of Transmission System Operators for Electricity~\cite{entsoe}. As time series data for hydrogen prices are not readily available, the hydrogen price time series are derived from electricity prices. First, the electricity prices are scaled to an average of 3~€/kg, then random variations of $\pm25\%$ are added to the scaled values. Finally, the resulting values are limited within the range of 1–5~€/kg (Figure~\ref{fig:Hydrogen price in 2017}). The wind capacity factors are bias-corrected reanalysis data~\cite{STAFFELL20161224} obtained from~\cite{RenewablesNinja}.

\section{Steady controller}
\label{S:Steady controller}

The steady controller is based on a constant $\bar{M}_{2}^d$ daily amount of hydrogen that should be sent to the HPA. This controller adjusts for the amount of hydrogen that could not be produced the previous day, correcting for drift from steady production. The drift occurs when wind resources are insufficient to produce the specified amount of hydrogen, necessitating a compensatory mechanism. The $\bar{M}_{2}^d$ is determined by Eqs.~(\ref{eq:steady_controller}) and~(\ref{eq:filter1})

\begin{equation}
\label{eq:steady_controller}
\widetilde{M}_{2}^d = d \times \frac{{M}_2^*}{T} - \sum_{i=0}^{d-1} M_2^i \quad \text{for} \quad 1 \leq d \leq T, \quad M_2^0 = 0
\end{equation}

\begin{equation}
\label{eq:filter1}
\overline{M}_{2}^d = \min\left(\widetilde{M}_2^d, \widehat{M}_1^d\right)
\end{equation}
where $T$ and $d$ are the period of the HPA and the number of days into the HPA respectively. The variables $\widetilde{M}_{2}^d$, ${M}_2^*$, and $M_2^i$ are the unrestricted hydrogen mass target for the HPA set by the controller for day \(d\), the total hydrogen mass target in the HPA, and the hydrogen mass sent to the HPA on day $i$ of the contract respectively. The condition $M_2^0 = 0$ means that all the hydrogen mass specified in the HPA should be delivered within the HPA period as no hydrogen is sent to the HPA before the beginning of the HPA period. The variable $\widehat{M}_{1}^d$ in Eq~\ref{eq:filter1} is the maximum hydrogen mass that can be produced on day \(d\) considering the available wind resources, irrespective of electricity prices and potential revenue. Equation~\ref{eq:filter1} ensures that $\bar{M}_{2}^d$ is within the production limits of the system. 

\nomenclature{$T$}{Period of the hydrogen purchase agreement (days)}
\nomenclature{$d$}{Number of days into the hydrogen purchase agreement}
\nomenclature{$\widetilde{M}_{2}^d$}{Hydrogen mass target for the hydrogen purchase agreement set by the controller for day \(d\)}
\nomenclature{${M}_2^*$}{Total hydrogen mass target for the hydrogen purchase agreement (kg)}
\nomenclature{$M_2^i$}{Hydrogen mass sent to the hydrogen purchase agreement on day {$i$} (kg)}
\nomenclature{$\overline{M}_{2}^d$}{Hydrogen mass target for the hydrogen purchase agreement on day \(d\) (kg)}
\nomenclature{$\widehat{M}_{1}^d$}{Maximum hydrogen mass that can be produced on day \(d\)}

\section{Cumulative hydrogen export to the HPA using the BFLC}

\begin{figure*}
    \centering
    \begin{subfigure}[b]{0.45\textwidth}
        \centering        \includegraphics[width=\linewidth]{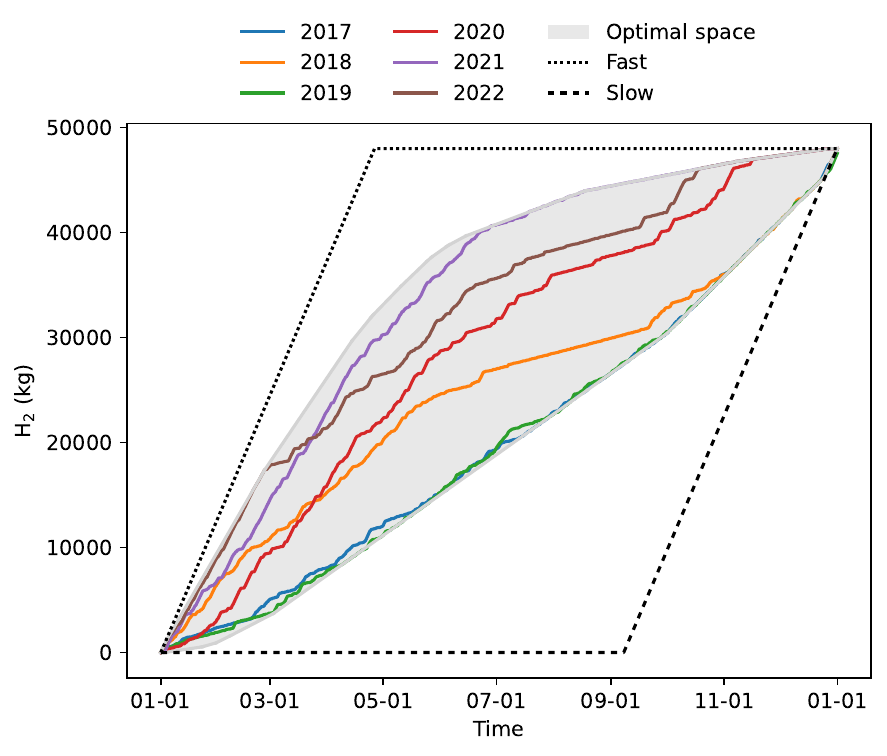}
        \caption{In-sample}
        \label{fig:Cumulative hydrogen export to the HPA using the BFLC (in sample)}
    \end{subfigure}
    \hfill
    \begin{subfigure}[b]{0.45\textwidth}
        \centering
        \includegraphics[ width=\linewidth]{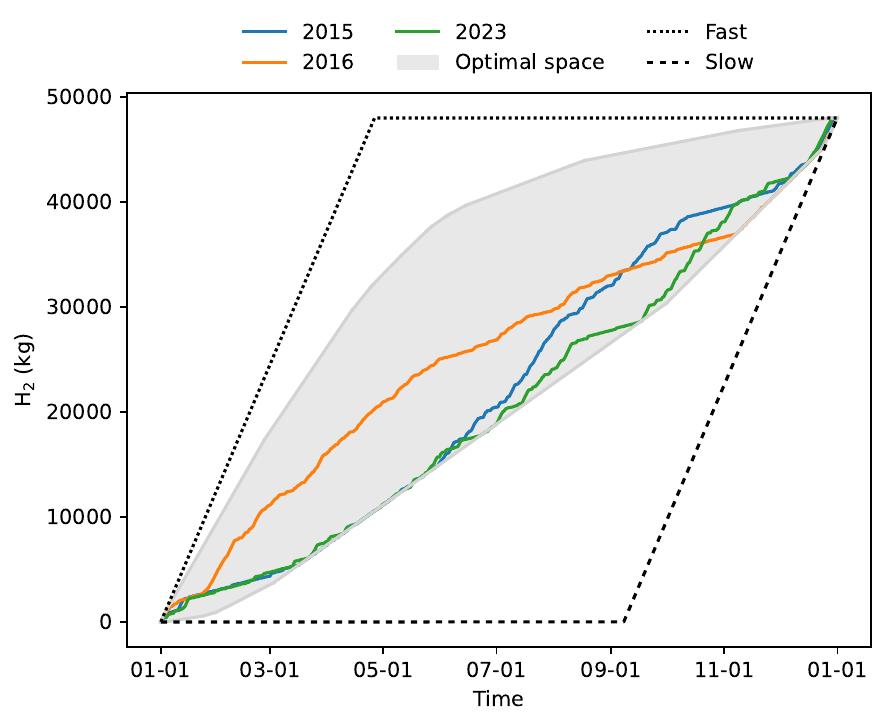}
        \caption{Out-of-sample}
        \label{fig:Cumulative hydrogen export to the HPA using the BFLC (out of sample)}
    \end{subfigure}
    \caption{Cumulative hydrogen export to the HPA under BFLC for (a) in-sample and (b) out-of-sample data. The grey-shaded region represents the optimal export space derived from the benchmark results, ensuring cumulative exports remain within predefined boundaries.}
    \label{fig:Cumulative hydrogen export to the HPA using the BFLC}
\end{figure*}

Figure~\ref{fig:Cumulative hydrogen export to the HPA using the BFLC} presents the cumulative hydrogen exports to the HPA when the daily export of hydrogen to the HPA is determined by the BFLC. The results underscore that the optimal space boundaries are essential to satisfy the delivery target of 48000~kg. In 2020–-2022, the upper boundary limits the daily exports, preventing the cumulative exports from exceeding the target. Conversely, in 2015-–2019 and 2023, the lower boundary enforces that hydrogen export to the HPA is not excessively slow, ensuring that the export target is reached by the end of the year.

\section{Revenue analysis for a United Kingdom case study}
\label{S:Revenue analysis for a United Kingdom case study}

\begin{figure}
    \centering
    \includegraphics[trim=0.5cm 0.0cm 0.0cm 0.5cm, clip, width=0.45\textwidth]{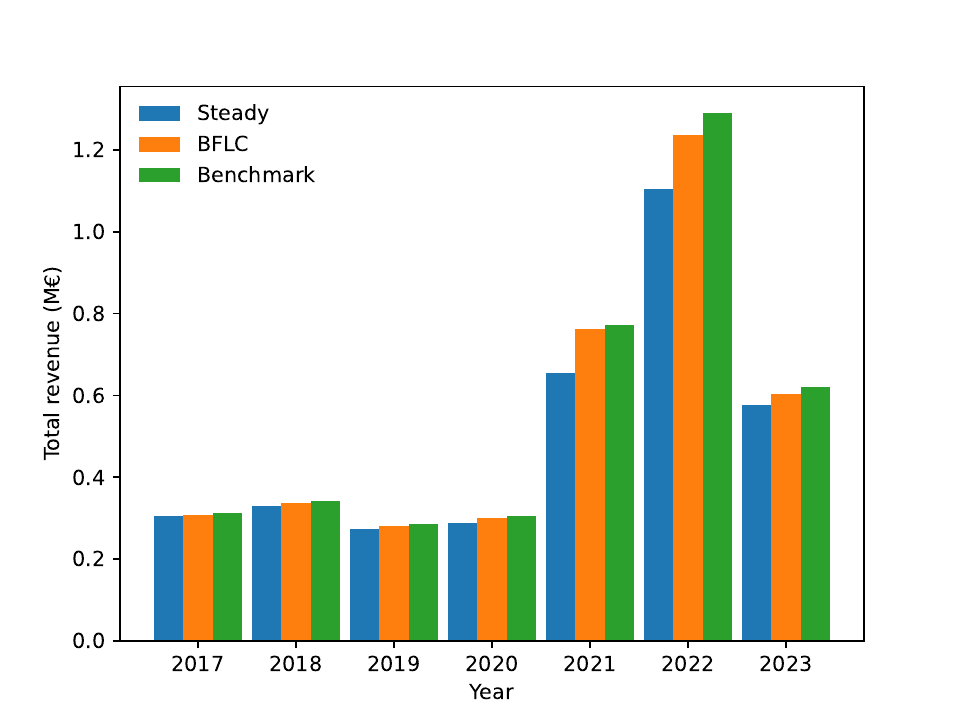}
    \caption{Comparison of total revenues achieved in the with the steady control (blue), BFLC (orange), and the benchmark approach (green) in the United Kingdom}
    \label{fig:Total revenue (United Kingdom)}
\end{figure}

\begin{figure}
    \centering
    \includegraphics[trim=0.5cm 0.0cm 0.0cm 0.0cm, clip, width=0.4\textwidth]{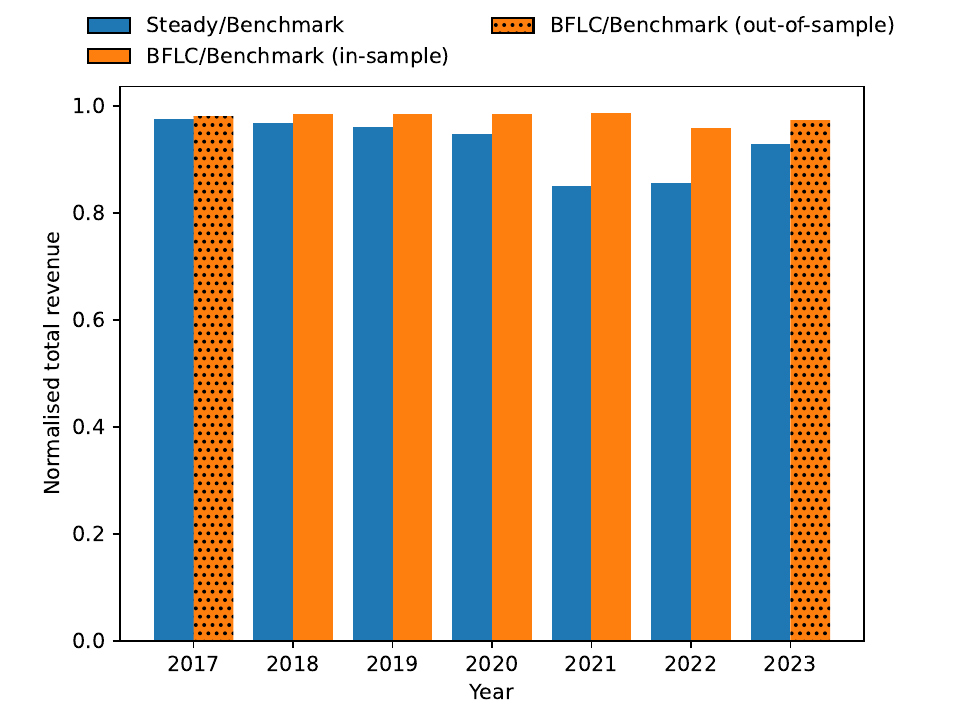}
    \caption{Comparison of normalised total revenues achieved with steady control (blue), in-sample BFLC (orange), and out-of-sample BFLC (orange, black dotted) in the United Kingdom}
    \label{fig:Normalised total revenue (United Kingdom)}
\end{figure}

The results in Figures~\ref{fig:Total revenue (United Kingdom)} and~\ref{fig:Normalised total revenue (United Kingdom)} are for a site located in the United Kingdom (latitude: 53.4883\degree, longitude: 0.1216\degree). These results are based on the same methodology described for the Danish case study.
The total revenues achieved by steady control and BFLC follow a pattern similar to that of the benchmark, with the BFLC consistently achieving higher revenues than the steady control. During periods of high electricity price levels and high variability, especially in 2021 and 2022, the normalised total revenue with steady control is significantly lower than those of the benchmark. During these two years, a larger revenue difference can be observed between the BFLC and the steady control, highlighting the benefits of the BFLC.

\bibliography{Bibliography}

\begin{thebibliography}{10}

\bibitem{deng2025global}
Zhu Deng, Biqing Zhu, Steven~J Davis, Philippe Ciais, Dabo Guan, Peng Gong, and Zhu Liu.
\newblock Global carbon emissions and decarbonization in 2024.
\newblock {\em Nature Reviews Earth \& Environment}, 6(4):231--233, 2025.

\bibitem{lee2023ipcc}
Hoesung Lee, Katherine Calvin, Dipak Dasgupta, Gerhard Krinner, Aditi Mukherji, Peter Thorne, Christopher Trisos, Jos{\'e} Romero, Paulina Aldunce, Ko~Barret, et~al.
\newblock Ipcc, 2023: Climate change 2023: Synthesis report, summary for policymakers. contribution of working groups i, ii and iii to the sixth assessment report of the intergovernmental panel on climate change [core writing team, h. lee and j. romero (eds.)]. ipcc, geneva, switzerland., 2023.

\bibitem{change2022mitigating}
Climate Change.
\newblock Mitigating climate change.
\newblock {\em Working Group III contribution to the sixth assessment report of the intergovernmental panel on climate change}, 2022.

\bibitem{iea2023co2}
{International Energy Agency}.
\newblock Co$_2$ emissions in 2022, 2023.

\bibitem{renewables2022}
IEA.
\newblock Renewables 2022.
\newblock \url{https://www.iea.org/reports/renewables-2022}, December 2022.
\newblock License:C BY 4.0.

\bibitem{irena2025}
{International Renewable Energy Agency (IRENA)}.
\newblock {Renewable Capacity Statistics 2025}, March 2025.
\newblock Accessed: 2025-05-15.

\bibitem{mathiesen2015smart}
Brian~Vad Mathiesen, Henrik Lund, David Connolly, Henrik Wenzel, Poul~Alberg {\O}stergaard, Bernd M{\"o}ller, Steffen Nielsen, Iva Ridjan, Peter Karn{\o}e, Karl Sperling, et~al.
\newblock Smart energy systems for coherent 100\% renewable energy and transport solutions.
\newblock {\em Applied energy}, 145:139--154, 2015.

\bibitem{cole2021cost}
Wesley Cole, A~Will Frazier, and Chad Augustine.
\newblock Cost projections for utility-scale battery storage: 2021 update.
\newblock Technical report, National Renewable Energy Lab.(NREL), Golden, CO (United States), 2021.

\bibitem{sivaram2018need}
Varun Sivaram, John~O Dabiri, and David~M Hart.
\newblock The need for continued innovation in solar, wind, and energy storage.
\newblock {\em Joule}, 2(9):1639--1642, 2018.

\bibitem{guerra2021beyond}
Omar~J Guerra.
\newblock Beyond short-duration energy storage.
\newblock {\em Nature Energy}, 6(5):460--461, 2021.

\bibitem{sterner2021power}
Michael Sterner and Michael Specht.
\newblock Power-to-gas and power-to-x—the history and results of developing a new storage concept.
\newblock {\em Energies}, 14(20):6594, 2021.

\bibitem{oliveira2021green}
Alexandra~M Oliveira, Rebecca~R Beswick, and Yushan Yan.
\newblock A green hydrogen economy for a renewable energy society.
\newblock {\em Current Opinion in Chemical Engineering}, 33:100701, 2021.

\bibitem{ball2009future}
Michael Ball and Martin Wietschel.
\newblock The future of hydrogen--opportunities and challenges.
\newblock {\em International journal of hydrogen energy}, 34(2):615--627, 2009.

\bibitem{zeyen2023endogenous}
Elisabeth Zeyen, Marta Victoria, and Tom Brown.
\newblock Endogenous learning for green hydrogen in a sector-coupled energy model for europe.
\newblock {\em Nature communications}, 14(1):3743, 2023.

\bibitem{neumann2023potential}
Fabian Neumann, Elisabeth Zeyen, Marta Victoria, and Tom Brown.
\newblock The potential role of a hydrogen network in europe.
\newblock {\em Joule}, 7(8):1793--1817, 2023.

\bibitem{westwood2025hydrogen}
Westwood~Global Energy.
\newblock Over a fifth of all european hydrogen projects stalled or cancelled, 2025.
\newblock Accessed May 2025.

\bibitem{orsted2024h1}
{\O}rsted A/S.
\newblock Interim report for the first half year of 2024 – increased earnings from offshore sites, progress on our business plan, and commissioning of around 2\,gw renewable capacity, August 2024.

\bibitem{reuters2024iberdrola}
Reuters.
\newblock Spain’s iberdrola slashes green hydrogen target.
\newblock {\em Reuters}, 2024.
\newblock Accessed May 2025.

\bibitem{aurora2024bankability}
{Aurora Energy Research}.
\newblock Comment: Bankability and hpas.
\newblock \url{https://auroraer.com/resources/aurora-insights/articles/comment-bankability-and-hpas}, 2024.
\newblock Accessed: 2025-06-30.

\bibitem{h2global2022}
{Timo Bollerhey, Markus Exenberger, Florian Geyer, Kirsten Westphal}.
\newblock H2global -- idea, instrument \& intentions.
\newblock Policy Brief 01/2022, H2Global Stiftung, 2022.
\newblock Available at: \url{https://files.h2-global.de/H2Global-Stiftung-Policy-Brief-01_2022-EN.pdf} (accessed: 2025-04-24).

\bibitem{bokde2020graphical}
Neeraj Bokde, Bo~Tranberg, and Gorm~Bruun Andresen.
\newblock {A graphical approach to carbon-efficient spot market scheduling for Power-to-X applications}.
\newblock {\em Energy Conversion and Management}, 224:113461, 2020.

\bibitem{FARAH_green_hydrogen}
Sleiman Farah, Neeraj Bokde, and Gorm~Bruun Andresen.
\newblock Cost and co2 emissions co-optimisation of green hydrogen production in a grid-connected renewable energy system.
\newblock {\em International Journal of Hydrogen Energy}, 84:164--176, 2024.

\bibitem{FARAH_green_hydrogen_conference}
Sleiman Farah and Gorm~B. Andresen.
\newblock Green hydrogen production: cost and co2 emissions co-optimisation.
\newblock In {\em 8th International Hybrid Power Plants \& Systems Workshop (HYB 2024)}, volume 2024, pages 172--176, 2024.

\bibitem{Soliman_Adaptive_Fuzzy_Logic_wind_turbine}
Mahmoud~A. Soliman, Hany~M. Hasanien, Haitham~Z. Azazi, E.~E. El-Kholy, and Sabry~A. Mahmoud.
\newblock An adaptive fuzzy logic control strategy for performance enhancement of a grid-connected pmsg-based wind turbine.
\newblock {\em IEEE Transactions on Industrial Informatics}, 15(6):3163--3173, 2019.

\bibitem{Belman-Flores_Fuzzy_Logic_Control_refrigeration_systems}
Juan~Manuel Belman-Flores, David~Alejandro Rodríguez-Valderrama, Sergio Ledesma, Juan~José García-Pabón, Donato Hernández, and Diana~Marcela Pardo-Cely.
\newblock A review on applications of fuzzy logic control for refrigeration systems.
\newblock {\em Applied Sciences}, 12(3), 2022.

\bibitem{Arifin_steering_Fuzzy_Logic_Control}
Bustanul Arifin, Bhakti~Yudho Suprapto, Sri Arttini~Dwi Prasetyowati, and Zainuddin Nawawi.
\newblock Steering control in electric power steering autonomous vehicle using type-2 fuzzy logic control and pi control.
\newblock {\em World Electric Vehicle Journal}, 13(3), 2022.

\bibitem{Liu_fuzzy_logic_control_hydrogen_production}
Ziqi Liu, Mohammed~J. Beshir, and Zhongxia Zhang.
\newblock A fuzzy logic controller design in an off-grid microgrid with hydrogen production.
\newblock In {\em 2024 7th International Conference on Electrical Engineering and Green Energy (CEEGE)}, pages 72--77, 2024.

\bibitem{Directorate_General_for_Energy}
{Directorate-General for Energy}.
\newblock {Delegated regulation on Union methodology for RFNBOs}.
\newblock \url{https://eur-lex.europa.eu/legal-content/EN/TXT/PDF/?uri=CELEX:32023R1184}, 2023.
\newblock Accessed on 19/May/2023.

\bibitem{danish_energy_agency2024}
{The Danish Energy Agency}.
\newblock {\em Technology Data for Renewable Fuels}, 2024.
\newblock Latest update: April 2024.

\bibitem{awoe_hydrogen_compression2022}
{A World Of Energy}.
\newblock {\em Hydrogen Compression}, 2022.
\newblock Published on January 16, 2022.

\bibitem{PyPSA}
T.~Brown, J.~H\"orsch, and D.~Schlachtberger.
\newblock {PyPSA: Python for Power System Analysis}.
\newblock {\em Journal of Open Research Software}, 6(4), 2018.

\bibitem{gurobi}
{Gurobi Optimization, LLC}.
\newblock {Gurobi Optimizer Reference Manual}.
\newblock \url{https://www.gurobi.com}, 2023.

\bibitem{warner2024scikitfuzzy}
Josh Warner, Jason Sexauer, Wouter~Van den Broeck, Bruno~P. Kinoshita, Jakub Balinski, et~al.
\newblock Jdwarner/scikit-fuzzy: Scikit-fuzzy 0.5.0, 2024.

\bibitem{ZADEH1965338}
L.A. Zadeh.
\newblock Fuzzy sets.
\newblock {\em Information and Control}, 8(3):338--353, 1965.

\bibitem{ross2005fuzzy}
Timothy~J Ross.
\newblock {\em Fuzzy logic with engineering applications}.
\newblock John Wiley \& Sons, 2005.

\bibitem{pyswarm}
Abraham Lee.
\newblock pyswarm 0.6.
\newblock \url{https://pypi.org/project/pyswarm/}, 2014.
\newblock Released: October 22, 2014; Accessed: 7 January 2025.

\bibitem{Generating_fuzzy_rules_by_learning_from_examples}
L.-X. Wang and J.M. Mendel.
\newblock Generating fuzzy rules by learning from examples.
\newblock {\em IEEE Transactions on Systems, Man, and Cybernetics}, 22(6):1414--1427, 1992.

\bibitem{entsoe}
ENTSO-E.
\newblock Day-ahead prices.
\newblock \url{https://newtransparency.entsoe.eu/}, 2024.
\newblock Accessed on 16/June/2025.

\bibitem{STAFFELL20161224}
Iain Staffell and Stefan Pfenninger.
\newblock Using bias-corrected reanalysis to simulate current and future wind power output.
\newblock {\em Energy}, 114:1224--1239, 2016.

\bibitem{RenewablesNinja}
Stefan Pfenninger and Iain Staffell.
\newblock Renewables.ninja.
\newblock \url{https://www.renewables.ninja/}, 2024.
\newblock Accessed: 16/June/2025.

\end{thebibliography}
\bibliographystyle{unsrt}

\end{document}